\documentclass[conference,10pt]{IEEEtran}
\usepackage{cite}
\usepackage{epsfig}
\usepackage{epstopdf}
\usepackage{graphicx}
\usepackage{xcolor}
\usepackage{tikz}
\usetikzlibrary{arrows,positioning,shapes.geometric,circuits.logic.US}
\usepackage{pgfplots}
\usepackage{multirow}
\usepackage{upgreek}
\usepackage{amssymb}
\usepackage{amsmath}
\usepackage{cases}
\usepackage{url}
\usepackage{pifont}
\usepackage{bm}
\usepackage{amsthm}
\usepackage{balance}
\usepackage{readarray}

\usepackage{tabularx}
\usepackage{booktabs}
\usepackage{filecontents}
\usepackage{subcaption}
\newcolumntype{Y}{>{\centering\arraybackslash}X}

\usepackage[linesnumbered, ruled, vlined, noend]{algorithm2e}
\usepackage{algpseudocode}
\usepackage{array}

\newcommand{\fixme}[2]{\ifx&#2&{\leavevmode\color{red}#1}\else{\leavevmode\color{red}FIXME\{}#1{\leavevmode\color{red}\}}\footnote{{\leavevmode\color{red}#2}}\PackageWarning{Fixme}{#1: #2}\fi}

\newcommand{\newstuff}[2]{\ifx&#2&{\leavevmode\color{blue}#1}\else{\leavevmode\color{blue}FIXME\{}#1{\leavevmode\color{blue}\}}\footnote{{\leavevmode\color{blue}#2}}\PackageWarning{Newstuff}{#1: #2}\fi}

\DeclareMathOperator*{\sgn}{sgn}

\begin{document}

\title{On List Decoding of 5G-NR Polar Codes}

\author{\IEEEauthorblockN{Charles Pillet, Valerio Bioglio, Carlo Condo}
\IEEEauthorblockA{\\Mathematical and Algorithmic Sciences Lab\\ France Research Center, Huawei Technologies Co. Ltd.\\
Email: $\{$charles.pillet1,valerio.bioglio,carlo.condo$\}$@huawei.com}}

\maketitle
\begin{abstract}
The 5$^\text{th}$ generation wireless systems (5G) standardization process of the 3$^\text{rd}$ generation partnership project (3GPP) chose polar codes as a channel coding scheme for the control channel.  
In case of downlink control information, polar codes are concatenated with distributed distributed cyclic redundancy check (CRC). 
Whereas CRC bits allow to improve the performance of successive cancellation list (SCL) decoders by improving distance properties, distributed CRC bits allow for path pruning and decoding early-termination. 
In this paper, we show how to take advantage of the distributed CRC to improve SCL decoding, analyzing various schemes having different early-termination and error correction properties. 
Simulation results compare the proposed decoding schemes, showing different trade-offs between error-correction performance and early-termination with different decoder parameters.
\end{abstract}


\IEEEpeerreviewmaketitle

\section{Introduction} \label{sec:intro}
Polar codes are a family of channel codes asymptotically achieving channel capacity under the low-complexity \emph{successive cancellation} (SC) decoder \cite{arikan}. 
However, polar codes under SC decoding have only mediocre error-correction performance at finite code length.
The \emph{successive cancellation list} (SCL) decoder proposed in \cite{tal_list} enhances SC by considering $L$ candidate codewords during the decoding, providing better block error rate (BLER) performance at the cost of a higher complexity. 
Moreover, a cyclic redundancy check (CRC) code can be appended to polar codes to further improve the BLER performance of an SCL decoder \cite{CRC_aid}; in this case, the decoder is referred to as \emph{CRC-aided SCL} (CA-SCL) decoder. 
Optimal CRC selection is of paramount importance to improve the error correction performance of CA-SCL \cite{Hashemi_5G}.

The 5$^\text{th}$ generation wireless systems (5G) standardization process of the 3$^\text{rd}$ generation partnership project (3GPP) chose polar codes as a channel coding scheme for the control channel of the enhanced mobile broad band scenario (eMBB) \cite{3GPP_TS}. 
CRC is considered in all use cases, but for the downlink control information, polar codes are concatenated with distributed CRC, whose bits are obtained by interleaving the bits between the CRC encoder and the polar encoder. 
The interleaver distributes the CRC bits such that a CRC bit is positioned after the last bit needed for its calculation \cite{hui_dca}. 
As a consequence, the decoding complexity can be reduced by early terminating the decoding if an incorrect check is encountered \cite{dist_CRC_aid,Blind_ET}, or alternatively error-correction can be improved by pruning the SCL decoding tree as a \emph{Parity-Check-Concatenated} (PCC) polar codes \cite{PC_conc}.

In this paper, we consider polar codes for the physical broadcast channel (PBCH) and physical downlink control channel (PDCCH) in 5G standard, and analyze different methods to exploit the information provided by the distributed CRC bits during the decoding. 
The decoder can either focus on early termination \cite{dist_CRC_aid,Blind_ET}, discard paths not passing the partial CRC check or use CRC bits as dynamic frozen bits \cite{PC_conc}.
For each decoding strategy, we provide an accurate description of the decoder and simulation results in terms of early termination and BLER performance.  
The advantages and drawbacks of each decoding approach as code and decoder parameters vary are discussed.

\section{Preliminaries} \label{sec:prel}
In this section, we review basic concepts on polar code design and decoding. 

\subsection{Polar codes}
Polar codes rely on the polarization properties of the Kronecker powers of kernel matrix $G_2 \triangleq \begin{bmatrix}
   1 & 0 \\
   1 & 1
\end{bmatrix}$,
defining the transformation matrix $G_N = G_2^{\otimes n}$, with $N=2^n$. 
Polarization effect creates $N$ virtual channels with different reliabilities, each one able to transmit a single bit $u_i$. 
In an $(N,K)$ polar code of length $N$ and dimension $K$, the message bits are allocated in the $K$ most reliable channels, termed as the information set $\mathcal{I}$, while the others $N-K$ channels, constituting the frozen set $\mathcal{F}$, are set to 0. 
In fact, the input vector $\mathbf{u} = \{u_0,u_1,\ldots,u_{N-1}\}$ is generated by assigning $u_i = 0$ if $i \in \mathcal{F}$, and storing information in the remaining entries. 
Codeword $\mathbf{x}$ is computed as
\begin{align}
\mathbf{x}=\mathbf{u}\mathbf{G_N}.
\end{align}

\subsection{Successive-Cancellation Decoding}\label{subsec:decoding}
\begin{figure}[t!]
  \centering
  \begin{tikzpicture}[scale=1.9, thick]
\newcommand\Triangle[1]{-- ++(0:2*#1) -- ++(120:2*#1) --cycle}
\newcommand\Square[1]{+(-#1,-#1) rectangle +(#1,#1)}

  \fill [gray, very thick] (0,0) circle [radius=.05];
  

  \fill [gray, very thick] (-1.0,-.5) circle [radius=.05]; 
  \fill [gray, very thick] (1,-.5) circle [radius=.05]; 

  \draw (-1.5,-1) circle [radius=.05];
  \fill [gray, very thick] (-.5,-1.0) circle [radius=.05];
 \fill [gray, very thick] (.5,-1) circle [radius=.05];
  \fill (1.5,-1) circle [radius=.05];

  \draw (-1.75,-1.5) circle [radius=.05];
  \draw (-1.25,-1.5) circle [radius=.05];
  \draw (-.75,-1.5) circle [radius=.05];
  \fill (-.25,-1.5) circle [radius=.05];
  \draw (.25,-1.5) circle [radius=.05];
  \fill (.75,-1.5) circle [radius=.05];
  \fill (1.25,-1.5) circle [radius=.05];
  \fill (1.75,-1.5) circle [radius=.05];

  \node at (-1.75,-1.7) {$\hat{u}_0$};
  \node at (-1.25,-1.7) {$\hat{u}_1$};
  \node at (-.75,-1.7) {$\hat{u}_2$};
  \node at (-.25,-1.7) {$\hat{u}_3$};
  \node at (.25,-1.7) {$\hat{u}_4$};
  \node at (.75,-1.7) {$\hat{u}_5$};
  \node at (1.25,-1.7) {$\hat{u}_6$};
  \node at (1.75,-1.7) {$\hat{u}_7$};

  \draw (0,-.05) -- (-1,-.45);
  \draw (0,-.05) -- (1,-.45);

  \draw (-1,-.55) -- (-1.5,-.95);
  \draw (-1,-.55) -- (-.5,-.95);
  \draw (1,-.55) -- (.5,-.95);
  \draw (1,-.55) -- (1.5,-.95);

  \draw (-1.5,-1.05) -- (-1.75,-1.45);
  \draw (-1.5,-1.05) -- (-1.25,-1.45);
  \draw (-.5,-1.05) -- (-.75,-1.45);
  \draw (-.5,-1.05) -- (-.25,-1.45);
  \draw (.5,-1.05) -- (.25,-1.45);
  \draw (.5,-1.05) -- (.75,-1.45);
  \draw (1.5,-1.05) -- (1.25,-1.45);
  \draw (1.5,-1.05) -- (1.75,-1.45);

  \draw [very thin,gray,dashed] (-2,0) -- (2,0);
  \draw [very thin,gray,dashed] (-2,-.5) -- (2,-.5);
  \draw [very thin,gray,dashed] (-2,-1) -- (2,-1);
  \draw [very thin,gray,dashed] (-2,-1.5) -- (2,-1.5);

  \node at (-2.25,0) {$t=3$};
  \node at (-2.25,-.5) {$t=2$};
  \node at (-2.25,-1) {$t=1$};
  \node at (-2.25,-1.5) {$t=0$};

  \draw [->] (-.12,-.05) -- (-1,-.4) node [above=-.1cm,midway,rotate=25] {$\bm{\alpha}$};
  \draw [->] (-.88,-.45) -- (0,-.1) node [below=-.1cm,midway,rotate=25] {$\bm{\beta}$};

  \draw [->] (-1.06,-.55) -- (-1.5,-.9) node [above=-.1cm,midway,rotate=40] {$\bm{\alpha}^{\ell}$};
  \draw [->] (-1.44,-.95) -- (-1.0,-0.6) node [below=-.1cm,midway,rotate=40] {$\bm{\beta}^{\ell}$};

  \draw [<-] (-.94,-.55) -- (-.5,-.9) node [above=-.1cm,midway,rotate=-40] {$\bm{\beta}^{\text{r}}$};
  \draw [<-] (-.56,-.95) -- (-0.975,-.625) node [below=-.1cm,midway,rotate=-40] {$\bm{\alpha}^{\text{r}}$};

\end{tikzpicture}
  \caption{SC decoder of an $(8,4)$ polar code, $\mathcal{F} = \{u_0,u_1,u_2,u_4\}$.}
  \label{fig:tree}
\end{figure}
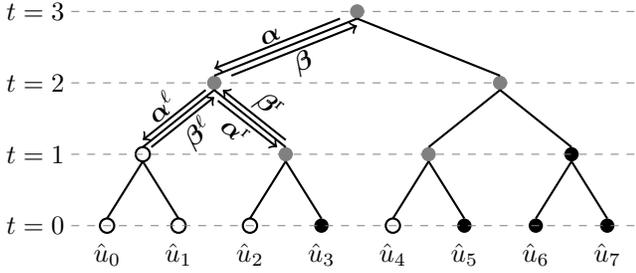

Polar codes were originally designed for SC decoding, as proposed in \cite{arikan}. 
The decoding process is portrayed in Figure~\ref{fig:tree} as a binary tree search with priority given to the left branch. 
At each tree stage $t$, the logarithmic likelihood ratios (LLRs) $\alpha$ are received from the parent node; $\alpha^l$ is computed and sent to the child node, and upon reception of the partial sums $\beta^l$, $\alpha^r$ can be computed. 
The partial sum vector $\beta$ is finally computed and sent to the parent node after receiving $\beta^r$. 
The LLR vectors $\alpha^l$ and $\alpha^r$ can be computed as 
\begin{align}
{\alpha}^l_j &= \sgn(\alpha_{j})\sgn(\alpha_{j+2^{t-1}}) \min(\alpha_{j},\alpha_{j+2^{t-1}}) \text{,} \label{eqn:alphaleft}\\
{\alpha}^r_j &= \alpha_{j+2^{t-1}} + (1-2\beta^{l}_{j})\alpha_{j} \text{,} \label{eqn:alpharight}
\end{align}
where $0 \leq j < 2^{t-1}$ and $\sgn(\cdot)$ returns the sign of the argument.
The partial sums are instead calculated as
\begin{equation}\label{eqn:beta}
  \beta_j=\left\{
  \begin{array}{@{}ll@{}}
    \beta^{l}_{j} \oplus \beta^{r}_{t}, & \text{if}~ j \leq 2^{t-1} \\
    \beta^{r}_{j-2^{t-1}}, & \text{otherwise,}
  \end{array}\right.
\end{equation} 
where $0 \leq j < 2^{t}$. 
When stage 0 is reached, the decoded bit $\hat{u}_i$ is equal to 
\begin{equation}
\hat{u}_i = \beta_{0} =
  \begin{cases}
    0 \text{,} & \text{if } i \in \mathcal{F} \text{ or } \alpha_{0}\geq 0\text{,}\\
    1 \text{,} & \text{otherwise,}
  \end{cases} \label{eq:decoded}
\end{equation}
where $0 \leq i < N$. 
This decoding approach, while being simple, suffers from error propagation due to its sequential nature. 
To mitigate this problem, SCL decoding has been proposed in \cite{tal_list} gathering $L=2^{m}$ parallel SC decoders with different decoding paths. 
At each information bit, the decoder considers both possible values, doubling the number of candidates and keeping the $L$ best ones according to a decision metric. 
This path metric can be computed as follows:
\begin{equation}
PM_{i} =
  \begin{cases}
    PM_{i-1}+ |\alpha^{i}_{0}| \text{,} & \text{if } \sgn\left(\alpha^{i}_{0}\right) \neq (-1)^{\hat{u}_i} \text{,}\\
    PM_{i-1} \text{,} & \text{otherwise.}
  \end{cases} \label{eq:pm}
\end{equation}
The path with the lowest $PM$ is chosen as the output of the SCL decoder.

\subsection{CRC-aided polar codes} \label{subsec:CA}
It has been shown in \cite{CRC_aid} that if a CRC code is appended to the information bits, the information provided by the CRC detector can be used as a codeword selection mechanism, substantially improving the decoder performance. 
The path relative to the smallest path metric fulfilling the CRC constraint is chosen as decoder output. 

In CA polar codes, $P$ CRC bits are computed from the $A$ information bits and appended at the end of the message, resulting in a vector $\mathbf{c}$ of size $K=A+P$. 
The CRC generator matrix \textbf{C} of size $A\times P$ can be constructed recursively through the CRC generator polynomial $g(x) = \sum_{k=0}^{P} = g_{k}x^{k}$ of degree $P$ as follows. 
The last row is given by the coefficients of $g(x)$ excluding the last one, i.e. $\mathbf{C}(A,i) = g_{P-i}$ with $i = \{1,..,P\}$, while previous row is computed as
\begin{align}
\mathbf{C}(k,i) &= \mathbf{C}(k+1,i+1) \oplus \left(\mathbf{C}(k+1,1) \times g_{P-i}\right)\\
\mathbf{C}(k,P) &= \mathbf{C}(k+1,1) \times g_{0}
\end{align}
Vector $\mathbf{c}$ can be computed as $\mathbf{c} = [\mathbf{a}|\mathbf{a} \times \mathbf{C}]$.

The $K$ bits obtained from the CRC encoder can be interleaved before being inserted into the input vector of the polar code \cite{hui_dca}. 
This interleaving is done to distribute the CRC bits equally inside the information bits while respecting that CRC remainder bit has to be encountered after its relevant information bits in the decoding process. 
This feature can be used by a distributed CA (DCA) decoder to reduce the decoding complexity by early terminating the decoding if every path is encountering an incorrect check. 
Moreover, the distributed CRC bits can be used to prune the SCL decoding tree and improve the decoder error-correction performance \cite{PC_conc}.

\subsection{DCA polar codes in 5G-NR}\label{subsec:dca}
\begin{figure}[t!]
  \centering
  \begin{tikzpicture}
\newcommand{\slashH}[3]{
\draw (#1-0.1,#2-0.1) node [below] {{#3}} -- (#1+0.1,#2+0.1) ;
}
\newcommand{\slashL}[3]{
\draw (#1+0.1,#2-0.1) node [below] {{#3}} -- (#1-0.1,#2+0.1) ;
}
\newcommand{\slashVD}[3]{
\draw (#1-0.1,#2+0.1) node [left] {{#3}} -- (#1+0.1,#2-0.1) ;
}
\newcommand{\slashVA}[3]{
\draw (#1-0.1,#2-0.1) node [left] {{#3}} -- (#1+0.1,#2+0.1) ;
}


\newcommand{\rlR}[8]{
\filldraw[draw=black,fill=#8] (#1,#2) rectangle (#1+#3,#2+#4) node[midway] {$\begin{tabular}{c} #5 \end{tabular}$};
\draw [->] (#1+#3, #2 + #4/2) node [above right] {#6} -- (#1+#3+#7,#2+#4/2);

}
\newcommand{\rlB}[8]{
\filldraw[draw=black,fill=#8] (#1,#2) rectangle (#1+#3,#2+#4) node[midway] {{#5}};

\draw [->] (#1+#3/2,#2) node [below right] {{#6}} -- (#1+#3/2,#2-#7);
}
\newcommand{\rlA}[8]{
\filldraw[draw=black,fill=#8] (#1,#2) rectangle (#1+#3,#2+#4) node[midway] {{#5}};
\draw [->] (#1+#3/2,#2+#4) node [above right] {{#6}} -- (#1+#3/2,#2+#4+#7);
}
\newcommand{\rlL}[8]{
\filldraw[draw=black,fill=#8] (#1,#2) rectangle (#1+#3,#2+#4) node[midway] {{#5}};
\draw [->] (#1, #2 + #4/2) node [above left] {{#6}} -- (#1-#7,#2+#4/2);
}
\def\Lrec{2.2}
\def\MiddleLrec{\Lrec/2}
\def\Wrec{1.2}
\def\MiddleWrec{\Wrec/2}
\def\Lfirstline{1}
\def\OutputLine{1}
\def\MiddleOutputLine{\OutputLine/2}
\def\MiddleO{0}
\def\UpperO{2}
\def\LowerO{-2}
\def\AbsLeftRec{\Lfirstline}
\def\AbsMiddleRec{\AbsLeftRec+\Lrec+\OutputLine}
\def\AbsRightRec{\AbsMiddleRec+\Lrec+\OutputLine}
\def\AbsRRR{\AbsRightRec+\Lrec+\OutputLine}

\slashH{\Lfirstline/2}{\MiddleO}{A}
\draw[->] (0,\MiddleO) node [above right] {\textbf{a}} -- (\AbsLeftRec, \MiddleO);
\rlR{\AbsLeftRec}{\MiddleO-\MiddleWrec}{\Lrec}{\Wrec}{}{\textbf{c}}{\OutputLine}{white}
\node (one) at (\AbsLeftRec+\MiddleLrec, \MiddleO) {\begin{tabular}{c} CRC \\ encoder \end{tabular}};
\slashH{\AbsLeftRec+\Lrec+\MiddleOutputLine}{\MiddleO}{K}
\filldraw[draw=black, fill=white] (\AbsMiddleRec,\MiddleO-\MiddleWrec) rectangle (\AbsMiddleRec+\Lrec,\MiddleO+\MiddleWrec) node[midway] {$\begin{tabular}{c} Interleaver \end{tabular}$};
\draw[->] (\AbsMiddleRec+\Lrec,\MiddleO) -- node [above] {\textbf{c'}} (\AbsMiddleRec+\Lrec+1, \MiddleO) -- (\AbsMiddleRec+\Lrec+1, \LowerO) -- (\AbsMiddleRec+\Lrec, \LowerO);
\slashVD{\AbsMiddleRec+\Lrec+\OutputLine}{\MiddleO/2+\LowerO/2}{K}
\filldraw[draw=black, fill=white] (\AbsMiddleRec,\LowerO-\MiddleWrec) rectangle (\AbsMiddleRec+\Lrec,\LowerO+\MiddleWrec);
\draw[->] (\AbsMiddleRec,\LowerO) -- (\AbsMiddleRec-\OutputLine, \LowerO);
\filldraw[draw=black, fill=white] (\AbsMiddleRec,\LowerO-\MiddleWrec) rectangle (\AbsMiddleRec+\Lrec,\LowerO+\MiddleWrec);
\draw[->] (\AbsMiddleRec,\LowerO) -- node [above right] {\textbf{u}}(\AbsMiddleRec-\OutputLine, \LowerO);
\slashL{\AbsMiddleRec-\OutputLine/2}{\LowerO}{N}
\node (one) at (\AbsMiddleRec+\MiddleLrec, \LowerO) {\begin{tabular}{c} Sub-channel \\ allocation \end{tabular}};
\filldraw[draw=black, fill=white] (\AbsLeftRec,\LowerO-\MiddleWrec) rectangle (\AbsLeftRec+\Lrec,\LowerO+\MiddleWrec);
\draw[->] (\AbsLeftRec,\LowerO) -- node [above right] {\textbf{x}} (\AbsLeftRec-\OutputLine, \LowerO);
\slashL{\AbsLeftRec-\OutputLine/2}{\LowerO}{N}
\node (one) at (\AbsLeftRec+\MiddleLrec, \LowerO) {\begin{tabular}{c} Polar code \\ encoder \end{tabular}};
\end{tikzpicture}
  \caption{DCA polar code encoding scheme.}
  \label{fig:dca_block}
\end{figure}
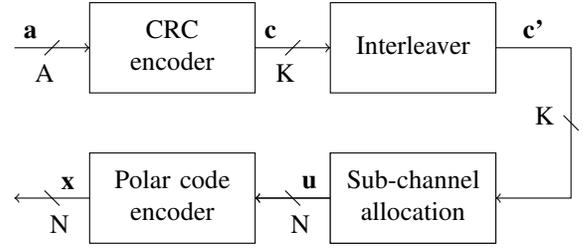

Polar codes were chosen as a coding scheme for the 5G new radio (5G-NR) for uplink and downlink control information (UCI and DCI) \cite{3GPP_TS}.
In this paper, we focus the discussion on the decoding process of the mother polar codes, without taking into account the rate matching strategy adopted \cite{design5G}. 
5G-NR polar codes include a distributed CRC for DCI, where the input bit interleaver is enabled for PBCH payloads and PDCCH DCIs.

Figure \ref{fig:dca_block} portrays the DCA polar code encoding scheme in 5G-NR.
The CRC polynomial adopted in the standard for DCI is 
\begin{align*}
g(x) =& x^{24}+x^{23}+x^{21}+x^{20}+x^{17}+x^{15}+x^{13}+x^{12}+ \\
        & +x^{8}+x^{4}+x^{2}+x+1.
\end{align*}

The interleaver input size is limited to $K \leq 164$, while its construction is calculated on the basis of the numeric sequence $\Pi_{IL}^{max}$ depicted in Table~\ref{tab:IB_inter} as follows. 
Parameter $h = 164 - K$ is calculated such that all the components of $\Pi_{IL}^{max}$ larger than $h$ are stored in the interleaver vector $\Pi$. 
Finally, $h$ is subtracted from all the entries of $\Pi$, such that $\Pi$ contains a set of indices from $0$ to $K-1$. 
The interleaving function is applied to $\mathbf{c}$, and the $K$-bit vector $\mathbf{c'} = \{ c_{\Pi(0)},\dots,c_{\Pi(K-1)} \}$ is obtained (Figure~\ref{fig:dca_block}).
We call $Q$ the subset of $\mathcal{I}$ containing the locations of the CRC bits. 
According to the presented construction, the number of CRC bits not located at the end of the message is variable and not larger than 8. 
\begin{table}[t!]
\begin{center}
\caption{Input bits interleaver pattern mother sequence (bold integers represent CRC bit indices).}
\label{tab:IB_inter}
\resizebox{0.48\textwidth}{!}{
\setlength{\extrarowheight}{1.7pt}
\begin{tabular}{|cccccccccccc|}
 \multicolumn{12}{c}{$\Pi_{IL}^{max}$} \\
 \hline
  0 &   2 &   4 &   7 &   9 &  14 &  19 &  20 &  24 &  25 &  26 &  28 \\
 31 &  34 &  42 &  45 &  49 &  50 &  51 &  53 &  54 &  56 &  58 &  59 \\
 61 &  62 &  65 &  66 &  67 &  69 &  70 &  71 &  72 &  76 &  77 &  81 \\
 82 &  83 &  87 &  88 &  89 &  91 &  93 &  95 &  98 & 101 & 104 & 106 \\
108 & 110 & 111 & 113 & 115 & 118 & 119 & 120 & 122 & 123 & 126 & 127 \\
129 & 132 & 134 & 138 & 139 & \textbf{140} &   1 &   3 &   5 &   8 &  10 &  15 \\
 21 &  27 &  29 &  32 &  35 &  43 &  46 &  52 &  55 &  57 &  60 &  63 \\
 68 &  73 &  78 &  84 &  90 &  92 &  94 &  96 &  99 & 102 & 105 & 107 \\
109 & 112 & 114 & 116 & 121 & 124 & 128 & 130 & 133 & 135 & \textbf{141} &   6 \\
 11 &  16 &  22 &  30 &  33 &  36 &  44 &  47 &  64 &  74 &  79 &  85 \\
 97 & 100 & 103 & 117 & 125 & 131 & 136 & \textbf{142} &  12 &  17 &  23 &  37 \\
 48 &  75 &  80 &  86 & 137 & \textbf{143} &  13 &  18 &  38 & \textbf{144} &  39 & \textbf{145} \\
 40 & \textbf{146} &  41 & \textbf{147} & \textbf{148} & \textbf{149} & \textbf{150} & \textbf{151} & \textbf{152} & \textbf{153} & \textbf{154} & \textbf{155} \\
\textbf{156} & \textbf{157} & \textbf{158} & \textbf{159} & \textbf{160} & \textbf{161} & \textbf{162} & \textbf{163}  & &  &  &  \\
\hline
\end{tabular}
}
\end{center}
\end{table}

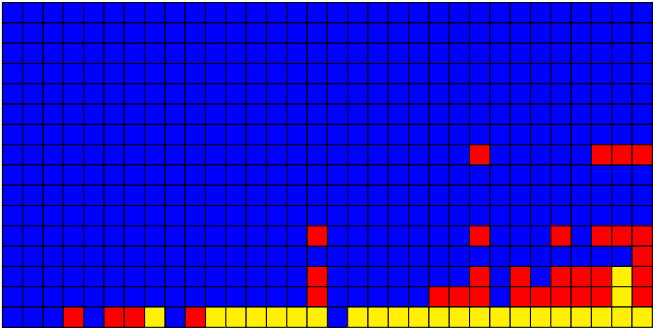
\begin{figure}[t!]
  \centering
 \begin{tikzpicture}
\def\data{
1 15 14 4
10 11 8 5
7 6 9 12
16 2 3 13
}

\def\Qfquo{0	0	0	0	0	0	0	0	0	0	0	0	0	0	0	0	0	0	0	0	0	0	0	0	0	0	0	0	0	0	0	0	1	1	1	1	1	1	1	1	1	1	1	1	1	1	1	1	1	1	1	1	1	1	1	1	1	1	1	1	1	1	1	1	2	2	2	2	2	2	2	2	2	2	2	2	2	2	2	2	2	2	2	2	2	2	2	2	2	2	2	2	2	2	2	2	3	3	3	3	3	3	3	3	3	3	3	3	3	3	3	3	3	3	3	3	3	3	3	3	3	3	3	3	3	3	3	3	4	4	4	4	4	4	4	4	4	4	4	4	4	4	4	4	4	4	4	4	4	4	4	4	4	4	4	4	4	4	4	4	5	5	5	5	5	5	5	5	5	5	5	5	5	5	5	5	5	5	5	5	5	5	5	5	5	5	5	5	5	5	5	5	6	6	6	6	6	6	6	6	6	6	6	6	6	6	6	6	6	6	6	6	6	6	6	6	6	6	6	6	6	6	6	6	7	7	7	7	7	7	7	7	7	7	7	7	7	7	7	7	7	7	7	7	7	7	7	7	7	7	7	7	8	8	8	8	8	8	8	8	8	8	8	8	8	8	8	8	8	8	8	8	8	8	8	8	8	8	8	8	8	8	8	8	9	9	9	9	9	9	9	9	9	9	9	9	9	9	9	9	9	9	9	9	9	9	9	9	9	9	9	9	9	9	9	9	10	10	10	10	10	10	10	10	10	10	10	10	10	10	10	10	10	10	10	10	10	10	10	10	10	10	10	10	10	10	10	10	11	11	11	11	11	11	11	11	11	11	11	11	11	11	11	11	11	11	11	11	11	11	11	11	11	11	12	12	12	12	12	12	12	12	12	12	12	12	12	12	12	12	12	12	12	12	12	12	12	12	12	12	12	12	12	12	12	13	13	13	13	13	13	13	13	13	13	13	13	13	13	13	13	13	13	13	13	13	13	13	13	14	14	14	14	14	14	14	14	14	14	14	14	14	14	14	14	14	14	14	14	14	15	15	15	15	15	15}
\def\Qfremainder{0 	1	2	3	4	5	6	7	8	9	10	11	12	13	14	15	16	17	18	19	20	21	22	23	24	25	26	27	28	29	30	31	0	1	2	3	4	5	6	7	8	9	10	11	12	13	14	15	16	17	18	19	20	21	22	23	24	25	26	27	28	29	30	31	0	1	2	3	4	5	6	7	8	9	10	11	12	13	14	15	16	17	18	19	20	21	22	23	24	25	26	27	28	29	30	31	0	1	2	3	4	5	6	7	8	9	10	11	12	13	14	15	16	17	18	19	20	21	22	23	24	25	26	27	28	29	30	31	0	1	2	3	4	5	6	7	8	9	10	11	12	13	14	15	16	17	18	19	20	21	22	23	24	25	26	27	28	29	30	31	0	1	2	3	4	5	6	7	8	9	10	11	12	13	14	15	16	17	18	19	20	21	22	23	24	25	26	27	28	29	30	31	0	1	2	3	4	5	6	7	8	9	10	11	12	13	14	15	16	17	18	19	20	21	22	23	24	25	26	27	28	29	30	31	0	1	2	3	4	5	6	7	8	9	10	11	12	13	14	15	16	17	18	19	20	21	22	24	25	26	27	28	0	1	2	3	4	5	6	7	8	9	10	11	12	13	14	15	16	17	18	19	20	21	22	23	24	25	26	27	28	29	30	31	0	1	2	3	4	5	6	7	8	9	10	11	12	13	14	15	16	17	18	19	20	21	22	23	24	25	26	27	28	29	30	31	0	1	2	3	4	5	6	7	8	9	10	11	12	13	14	15	16	17	18	19	20	21	22	23	24	25	26	27	28	29	30	31	0	1	2	3	4	5	6	7	8	9	10	11	12	13	14	16	17	18	19	20	21	22	24	25	26	28	0	1	2	3	4	5	6	7	8	9	10	11	12	13	14	15	16	17	18	19	20	21	22	23	24	25	26	27	28	29	30	0	1	2	3	4	5	6	7	8	9	10	11	12	13	14	16	17	18	19	20	21	22	24	26	0	1	2	3	4	5	6	7	8	9	10	11	12	13	14	16	17	18	19	20	24	0	1	2	4	8	16}
\def\Qiquo{7	7	7	7	11	11	11	11	11	11	12	13	13	13	13	13	13	13	13	14	14	14	14	14	14	14	14	14	14	14	15	15	15	15	15	15	15	15	15	15	15	15	15	15	15	15	15	15	15	15	15	15	15	15	15	15}
\def\Qiremainder{23	29	30	31	15	23	27	29	30	31	31	15	23	25	27	28	29	30	31	15	21	22	23	25	26	27	28	29	30	31	3	5	6	7	9	10	11	12	13	14	15	17	18	19	20	21	22	23	24	25	26	27	28	29	30	31}
\def\Qcrcquo{13	14	15	15	15	15	15	15	15	15	15	15	15	15	15	15	15	15	15	15	15	15	15	15}
\def\Qcrcremainder{30	30	7	10	11	12	13	14	15	17	18	19	20	21	22	23	24	25	26	27	28	29	30	31}
\readarray\Qfquo\dataFquo[456,1]
\readarray\Qfremainder\dataFR[456,1]
\readarray\Qiquo\dataIquo[56,1]
\readarray\Qiremainder\dataIR[56,1]
\readarray\Qcrcquo\dataCRCquo[24,1]
\readarray\Qcrcremainder\dataCRCR[24,1]

\def\stepsize{0.27}
\def\Lwidth{0.11mm}
\def\nbrows{16}
\def\nbcols{32}
\def\top{\nbrows*\stepsize}

\foreach \i in {1,...,\dataFquoROWS}{
           \filldraw[fill=blue] (\dataFR[\i,1]*\stepsize,\top-\dataFquo[\i,1]*\stepsize) rectangle (\dataFR[\i,1]*\stepsize+\stepsize,\top-\dataFquo[\i,1]*\stepsize-\stepsize);
}

\foreach \i in {1,...,\dataIRROWS}{
           \filldraw[fill=red] (\dataIR[\i,1]*\stepsize,\top-\dataIquo[\i,1]*\stepsize) rectangle (\dataIR[\i,1]*\stepsize+\stepsize,\top-\dataIquo[\i,1]*\stepsize-\stepsize);
}

\foreach \i in {1,...,\dataCRCquoROWS}{
           \filldraw[fill=yellow] (\dataCRCR[\i,1]*\stepsize,\top-\dataCRCquo[\i,1]*\stepsize) rectangle (\dataCRCR[\i,1]*\stepsize+\stepsize,\top-\dataCRCquo[\i,1]*\stepsize-\stepsize);
}
\end{tikzpicture}
  \caption{Example of subchannel allocation in PBCH frame.}
  \label{fig:PBCHframes}
\end{figure}

As an example, we describe the encoding of polar codes for PBCH in more detail.
The PBCH payload is mandatory composed of $A=32$ information bits and $P=24$ CRC bits, which requires an interleaver of size $K=56$, while the block length is $N=512$. 
The input vector of the obtained $(512,56)$ polar code is depicted in Figure~\ref{fig:PBCHframes}, where the blue, red and yellow squares represent frozen, information and CRC bits respectively.
In this case, only 3 CRC bits are actually distributed, while the position of the others has not been modified by the interleaver since they are located at the end of the input vector. 
After these steps, the polar encoded vector is interleaved and rate-matching is performed to obtain a codeword of length $E$ \cite{design5G}. 
In case of PBCH, the only admitted codeword length is $E=864$.

\section{5G-NR polar codes decoding} \label{sec:DCAdec}
In the following, three different incarnations of SCL decoding for DCA polar codes are presented, namely \emph{check and keep} (CK), \emph{check and remove} (CR) and \emph{check and select} (CS), each providing a different approach in the decision of which path to keep when facing a CRC bit. 

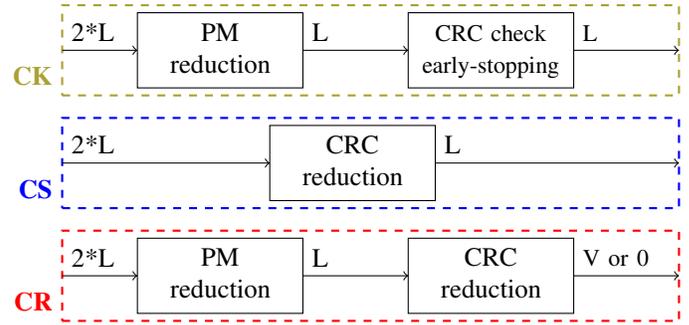
\begin{figure}[t!]
\centering
\begin{tikzpicture}

\newcommand{\slashH}[3]{
\draw (#1-0.1,#2-0.1) node [below] {{#3}} -- (#1+0.1,#2+0.1) ;
}
\newcommand{\slashVD}[3]{
\draw (#1-0.1,#2+0.1) node [left] {{#3}} -- (#1+0.1,#2-0.1) ;
}
\newcommand{\slashVA}[3]{
\draw (#1-0.1,#2-0.1) node [left] {{#3}} -- (#1+0.1,#2+0.1) ;
}
\newcommand{\xortikz}[3]{
\draw (#1,#2) circle(#3cm);
\draw (#1-#3,#2) -- (#1+#3,#2);
\draw (#1,#2-#3) -- (#1,#2+#3);
}

\newcommand{\rlR}[8]{
\filldraw[draw=black,fill=#8] (#1,#2) rectangle (#1+#3,#2+#4) node[midway] {$\begin{tabular}{c} #5 \end{tabular}$};
\draw [->] (#1+#3, #2 + #4/2) node [above right] {#6} -- (#1+#3+#7,#2+#4/2);

}
\newcommand{\rlB}[8]{
\filldraw[draw=black,fill=#8] (#1,#2) rectangle (#1+#3,#2+#4) node[midway] {{#5}};

\draw [->] (#1+#3/2,#2) node [below right] {{#6}} -- (#1+#3/2,#2-#7);
}
\newcommand{\rlA}[8]{
\filldraw[draw=black,fill=#8] (#1,#2) rectangle (#1+#3,#2+#4) node[midway] {{#5}};
\draw [->] (#1+#3/2,#2+#4) node [above right] {{#6}} -- (#1+#3/2,#2+#4+#7);
}
\newcommand{\rlL}[8]{
\filldraw[draw=black,fill=#8] (#1,#2) rectangle (#1+#3,#2+#4) node[midway] {{#5}};
\draw [->] (#1, #2 + #4/2) node [above left] {{#6}} -- (#1-#7,#2+#4/2);
}


\def\Lrec{2.2}
\def\MiddleLrec{\Lrec/2}
\def\Wrec{1}
\def\MiddleWrec{\Wrec/2}
\def\Lfirstline{1}
\def\OutputLine{1.4}
\def\MiddleOutputLine{\OutputLine/2}

\def\mid{0.8}
\def\MiddleO{0}
\def\UpperO{1.5}
\def\LowerO{-1.5}
\def\AbsLeftRec{\Lfirstline}
\def\AbsMiddleRec{\AbsLeftRec+\Lrec+\OutputLine}
\def\AbsRightRec{\AbsMiddleRec+\Lrec+\OutputLine}
\def\AbsRRR{\AbsRightRec+\Lrec+\OutputLine}
\draw [dashed, olive!80, thick] (0,\UpperO-\MiddleWrec*1.2) node[above left]{\textbf{CK}} rectangle (\AbsMiddleRec+\Lrec+\OutputLine, \UpperO+\MiddleWrec*1.2);
\draw [dashed, blue, thick] (0,\MiddleO-\MiddleWrec*1.2) node[above left]{\textbf{CS}} rectangle (\AbsMiddleRec+\Lrec+\OutputLine, \MiddleO+\MiddleWrec*1.2);
\draw [dashed, red, thick] (0,\LowerO-\MiddleWrec*1.2) node[above left]{\textbf{CR}} rectangle (\AbsMiddleRec+\Lrec+\OutputLine, \LowerO+\MiddleWrec*1.2);
\draw[->] (0,\LowerO)  node [above right] {2*L} -- (\AbsLeftRec, \LowerO);
\draw[->] (0,\MiddleO) node [above right] {2*L} -- (\AbsLeftRec+\mid*\Lrec, \MiddleO);
\draw[->] (0,\UpperO) node [above right] {2*L} -- (\AbsLeftRec, \UpperO);
\rlR{\AbsLeftRec}{\LowerO-\MiddleWrec}{\Lrec}{\Wrec}{}{L}{\OutputLine}{white}
\node (one) at (\AbsLeftRec+\MiddleLrec, \LowerO) {\begin{tabular}{c} PM \\ reduction \end{tabular}};
\rlR{\AbsLeftRec+\mid*\Lrec}{\MiddleO-\MiddleWrec}{\Lrec}{\Wrec}{}{L}{\OutputLine+\OutputLine+\Lrec-\mid*\Lrec}{white}
\node (one) at (\AbsLeftRec+\mid*\Lrec+\MiddleLrec, \MiddleO) {\begin{tabular}{c} CRC \\ reduction \end{tabular}};
\rlR{\AbsLeftRec}{\UpperO-\MiddleWrec}{\Lrec}{\Wrec}{}{L}{\OutputLine}{white}
\node (two) at (\AbsLeftRec+\MiddleLrec, \UpperO) {\begin{tabular}{c} PM \\ reduction \end{tabular}};
\rlR{\AbsMiddleRec}{\UpperO-\MiddleWrec}{\Lrec}{\Wrec}{}{\small{L}}{\OutputLine}{white}
\rlR{\AbsMiddleRec}{\LowerO-\MiddleWrec}{\Lrec}{\Wrec}{}{\small{V or 0}}{\OutputLine}{white}
\node (two) at (\AbsMiddleRec+\MiddleLrec, \LowerO) {\begin{tabular}{c} CRC \\ reduction  \end{tabular}};
\node (two) at (\AbsMiddleRec+\MiddleLrec, \UpperO) {\begin{tabular}{c} \small{CRC check}\\ \small{early-stopping} \end{tabular}};
\end{tikzpicture}
\caption{Behavior of CK, CS, CR decoders for CRC bits.}
\label{fig:crr}
\end{figure}

Part of these decoders were initially considered in preliminary works on the design and use of distributed CRC for SCL \cite{hui_dca,dist_CRC_aid,PC_conc}.
They are summarized in Figure \ref{fig:crr}, while Algorithm~\ref{alg:SCL} presents a high-level description of the standard SCL decoder with colored lines to highlight the steps performed by the different decoders. 
Blue lines (Lines 5 to 7) are performed only by CS, yellow lines (10 to 14) are performed by both the CK and CR decoders while orange lines (15 and 16) belong to the CR decoder exclusively.

\subsection{Check and Keep SCL decoder} \label{subsec:SCLCK}
In the CK SCL decoder, after every bit estimation the CRC of each path is computed. 
In case none of the $L$ surviving paths has a valid CRC, the decoding is terminated early. 
However, in case at least one of the CRC checks passes, all paths are kept.
At the end of the decoding, the output of the decoder is the path with the lowest $PM$ among those with a valid CRC. \\
The CK SCL decoder allows a simple, straightforward implementation based on the standard SCL decoder concept.  
The addition of an early termination criterion requires minimal modifications,  similar to those proposed within the blind detection framework in \cite{Blind_ET}.

\subsection{Check and Remove SCL decoder} \label{subsec:SCLCR}
The CR SCL decoder fully exploits the information provided by the distributed CRC, not only as an early-termination criteria. 
After every bit estimation, only the paths passing the CRC check are maintained, while the other ones are removed from the decoder. 
As a result, the number $V \leq L$ of surviving paths is not forced to be a power of two. 
As with the CK case, decoding is terminated early if no path passes the CRC check. 

The error-correction performance of CR SCL is expected to be better than that of CK, as failed path are removed from the decoder earlier, thus making room for paths that are likelier to be correct. 
For the same reason, the percentage of early terminations is lower in CR than in CK, as the list of candidates is allowed to grow only from paths that pass the CRC. 
The combined performance improvement and early termination capability comes at the implementation cost of having to manage a variable, number of parallel paths at any moment during the decoding.

 \begin{algorithm}[t!]
\def\HiLi{\leavevmode\rlap{\hbox to \hsize{\color{yellow!50}\leaders\hrule height .8\baselineskip depth .5ex\hfill}}}
\def\HiLiO{\leavevmode\rlap{\hbox to \hsize{\color{orange!50}\leaders\hrule height .8\baselineskip depth .5ex\hfill}}}
\def\HiLa{\leavevmode\rlap{\hbox to \hsize{\color{blue!30}\leaders\hrule height .8\baselineskip depth .5ex\hfill}}}
		\SetKwInOut{Input}{input}%
		\SetKwInOut{Output}{output}%
		\SetKw{Return}{return}
		\SetAlgoLined
		\Input{$N$ channel LLRs, list size $L$}
		\Output{$L$ vectors $\mathbf{\hat{u}}$, $L$ metrics}
		Initialize \textbf{A}, PMs, \textbf{B}, $\mathbf{\hat{u}}$ (Section~\ref{subsec:decoding})\;
		\For{$i = 0 \dots N-1$}{
		LLR update\;
		\uIf{$i\in \mathbf{I}$}{
		\HiLa\If{$i\in \mathbf{Q}$}{
		\HiLa Extend paths according to CRC\;
		\HiLa Go to Line \ref{alg:line_B}\;
		}
		Duplicate paths and update $PM_{i}$\;
		Keep the $L$ better paths ($u_i$ calculation)\;
		\HiLi update the $L$ CRC checksum\;
		\HiLi\If{$i \in \mathbf{Q}$}{
		\HiLi Compute the number of valid paths $V$\;
		\HiLi \If{$V=0$}{
		\HiLi break\;
		}
		\HiLiO Remove the $L-V$ wrong paths\;\label{alg:remove}
		\HiLiO Update current list size to $V$\;
		}
		}
		\Else{
		update PMs with $u_i = 0$\;
		}
		Update \textbf{B} for the $L$ lists\;\label{alg:line_B}
		}
		\Return{{\normalfont vector} $\mathbf{\hat{u}}$ {\normalfont with best metric} $PM_{i}$}
	\caption{CK/CR/CS SCL decoder}\label{alg:SCL}
\end{algorithm}
\subsection{Check and Select SCL decoder} \label{subsec:CS}
The CS SCL decoder uses the distributed CRC bits as dynamic frozen bits, or parity-check bits \cite{PC_conc}. 
When a CRC bit is reached, the SCL decoder treats it as an information bit, considering both its possible values and duplicating the number of paths.
After duplication, $L$ paths are thus passing the CRC while $L$ are not, and the latter are discarded.

The CS decoder takes a different approach than CR in how the CRC bits are treated: while in CR the LLR relative to the bit is allowed to contribute to the $PM$ calculation regardless of the check being passed or not, CS maximizes the chances that one of the $L$ decoders will return the correct information vector.
However, this approach prevents early-termination, resulting in a higher average decoding latency compared to the other two DCA-SCL incarnations.

\section{Simulation results} \label{sec:simu}
In the following, we compare the error-correction performance of the CK, CR and CS decoders.  
For fair comparison, each simulated frame is decoded by all three decoders.
Simulation results assume additive white Gaussian noise (AWGN) and binary phase-shift keying (BPSK) modulation and the error-correction performance is compared to the $\mathcal{O}\left(n^{-2}\right)$ approximation to Polyanskiy-Poor-Verdù (PPV) meta-converse bound \cite{SpectreMatlab}. The bound considers finite-block-length and an achievable probability of block error in order to provide an upper bound on the SNR.

\begin{figure}[t!]
\centering
\includegraphics[width=\columnwidth]{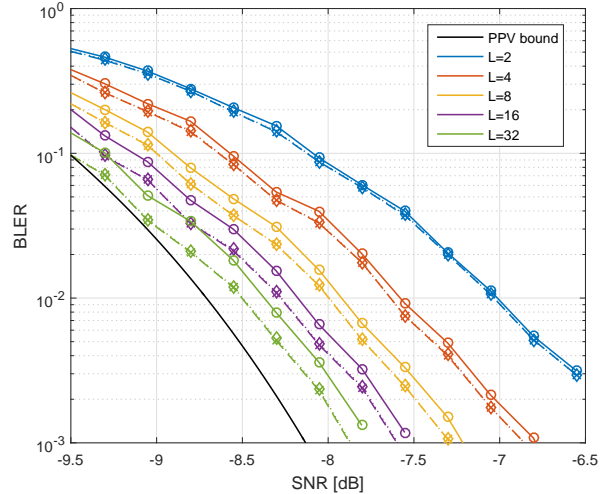}
\caption{BLER curves with the three decoders CK (solid line with circle markers), CR (dotted line with diamond markers) and CS (dashed line with cross markers) for PBCH and various list sizes.}
\label{fig:PBCHsimu}
\end{figure}

For the PBCH scenario, we consider code PC(512,56) having 24 CRC bits, 3 of which are distributed (Section~\ref{subsec:dca}).
Figure~\ref{fig:PBCHsimu} shows the BLER curves of the three decoders, obtained with different list sizes $L$. 
While at small list sizes the CK and CR decoders have similar BLER, it can be seen that as $L$ increases, the CR decoder has better and better error correction performance. 
With $L=8$, i.e. the baseline for 5G FEC performance evaluation \cite{design5G}, the gain is around $0.08$dB, while it reaches $0.2$dB for $L=32$.
Concerning the CS decoder, it grants a slight performance improvement with respect to the CR decoder for all list sizes, however without providing any early-termination.

\begin{figure}[t!]
\centering
\includegraphics[width=1.02\columnwidth]{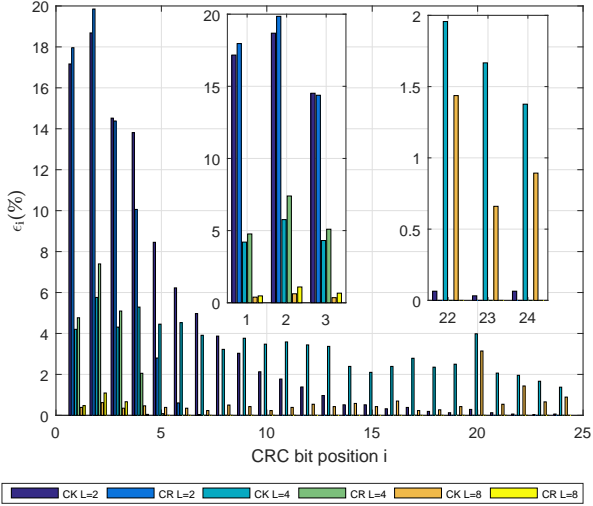}
\caption{Percentage of early-terminations for PC(512,56) with respect to distributed CRC bit.}
\label{fig:PBCHet}
\end{figure}

In order to explain Figure~\ref{fig:PBCHet} and Table~\ref{tab:PBCHet}, we introduce the following notations. 
We calculate the number $E_{tot}$ of failed decoding attempts as $E_{tot}=E_{e}+E_{w}$, where $E_{e}$ is the number of early-terminations, and $E_{w}$ is the the number of wrongly estimated codewords that went undetected by the CRC. 
Denoting as $E_{e}(i)$ the number of early-terminations at the $i^{th}$ CRC bit, then $E_{e}=\sum_{i=1}^{P}E_{e}(i)$.
Figure~\ref{fig:PBCHet} shows the percentage $\epsilon_i=E_e(i)/E$ of early terminations that occurred during PBCH simulation for both CK and CR given the CRC bit index.
The number of early terminations at the first CRC bit is the same for both decoders, as the CR decoder has not reached Line~\ref{alg:remove} in Algorithm~\ref{alg:SCL}; $\epsilon$ is smaller for the CK decoder because of its higher $E_{tot}$.
It can be observed that with small $L$, early-termination is more likely to happen, since the lower number of parallel paths increases the chance of having errors in all of them. 
Figure~\ref{fig:PBCHet} also shows that while CK maintains a noticeable early termination rate throughout all the CRC bit positions, the CR decoder will rarely early-stop once it reaches the undistributed CRC bits; for $L=2$, $\sum_{i=5}^{P}\epsilon_i=3.44\%$ compared to a total $\epsilon = \sum_{i=1}^{P}\epsilon_i=65.69\%$.

After decoding the last information bit, the CR decoder can early-stop only in case $L=V$ and the $L$ paths with a valid CRC are the ones associated to the $L$ highest path metrics after duplication.
If $L< 2V<2L$, the number of remaining paths will be between $L-V$ and $V$, and thus prevent early-termination.
If $2V\leq L$, no path metric reduction is performed, $V$ paths will pass the CRC while $V$ will not and be removed, thus maintaining the number of active paths constant.

Table~\ref{tab:PBCHet} shows the total percentage of early-terminations in case of failed decoding $\epsilon$ for code PC(512,56). 
Early-termination is less likely in the CR decoder than in the CK decoder; this is reflected in the improved BLER of CR.
Moreover, the percentage of early-terminations decreases very fast with the list size $L$. 
As a consequence, for large list sizes it can be preferable to use either the CS or the CR decoder.
\begin{table}[t!]
\begin{center}
\caption{Probability of early-termination for one frame error - PBCH.}
\label{tab:PBCHet}
\begin{tabular}{|c|c|c|c|c|c|}
 \hline
$L$ &   2 &   4 &   8 & 16 & 32\\
\hline
CK &  99.65 \% &  78.91 \% & 14.64 \% & 0.72 \% & 0.00 \%\\
\hline
CR & 65.69 \% & 19.40 \% & 2.28 \% & 0.13 \% & 0.00 \%\\
\hline
\end{tabular}
\end{center}
\end{table}

\begin{figure}[t!]
\centering
\includegraphics[width=1.05\columnwidth]{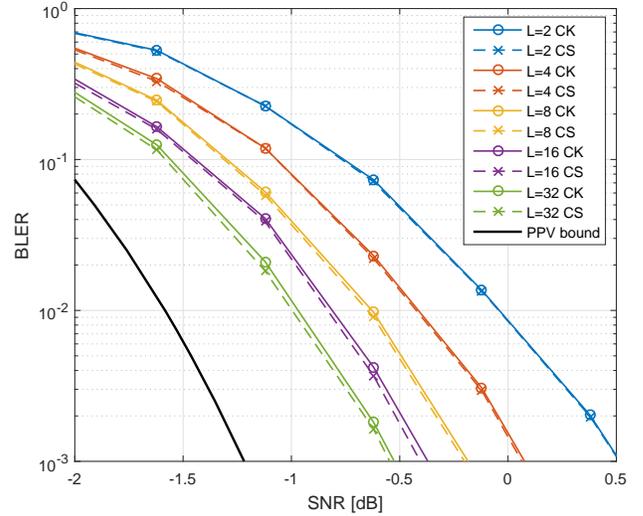}
\caption{PDCCH BLER curves for PC(512,164).}
\label{fig:PDCCHsimu}
\end{figure}

\begin{figure}[t!]
\centering
\includegraphics[width=1.05\columnwidth]{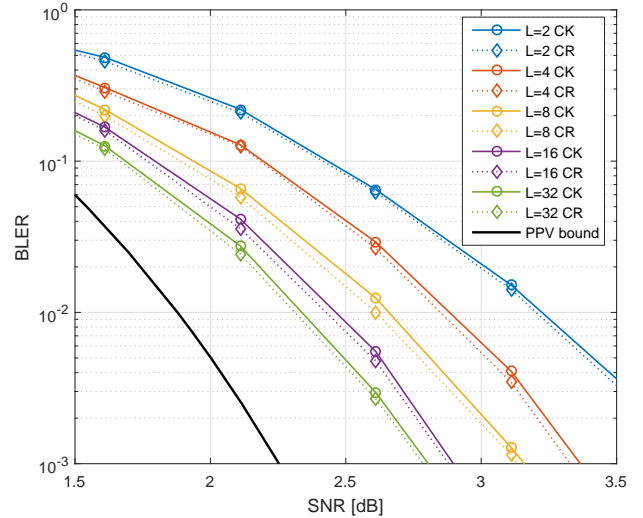}
\caption{PDCCH BLER curves for PC(256,152).}
\label{fig:PDCCHsimu7}
\end{figure}

Simulations have been run for the PDCCH case as well, using the same CRC polynomial as PBCH. In contrast to PBCH, PDCCH supports various message lengths $A$ and codeword lengths $E$.
Detailed results including early-termination of CK and CR decoders (Table~\ref{tab:PDCCHet}) and BLER curves are reported for PC(512,164) (Figure~\ref{fig:PDCCHsimu}) and for PC(256,152) (Figure~\ref{fig:PDCCHsimu7}).
In both cases, the codeword has the highest number of distributed CRC bits possible according to 5G standard, i.e. 7.
\begin{table}[t!]
\begin{center}
\caption{Probability of early-termination for one frame error - PDCCH.}
\label{tab:PDCCHet}
\begin{tabular}{|c|c|c|c|c|c|}
 \hline
$L$ &   2 &   4 &   8 & 16 & 32\\
\hline
\scriptsize{CK (512, 156)} &  99.87 \% &  81.35 \% & 18.75 \% & 2.02 \% & 0.80 \%\\
\hline
\scriptsize{CR (512, 156)} & 80.17 \% & 44.06 \% & 14.98 \% & 2.95 \% & 0.87 \%\\
\hline
\scriptsize{CK (256, 152)} &  99.87 \% &  79.66 \% & 14.88 \% & 0.71 \% & 0.24 \%\\
\hline
\scriptsize{CR (256, 152)} & 83.40 \% & 45.99 \% & 11.13 \% & 2.09 \% & 0.25 \%\\
\hline
\end{tabular}
\end{center}
\end{table}
Table \ref{tab:PDCCHet} shows that by having more distributed CRC bits, the CR decoder early-terminates more often compared to the PBCH scenario in case of failed decoding (Table \ref{tab:PBCHet}). 
Moreover, for large list size, the CR decoder early-terminates as often as CK decoder, a behavior not seen in PBCH. 
Furthermore, it can be noticed that the frequency of early-termination is not correlated with the rate of the 5G polar code used. 
The error-correction performance curves shown in Figure~\ref{fig:PDCCHsimu}-\ref{fig:PDCCHsimu7} depict the same behavior as PBCH, with CR and outperforming CK decoder. For the sake of clarity, only CR or CS was shown with CK, the decoders having similar performances.
A more noticeable gain can be observed at small list sizes with $N=256$.
\begin{figure}[t!]
\centering
\includegraphics[width=1.05\columnwidth]{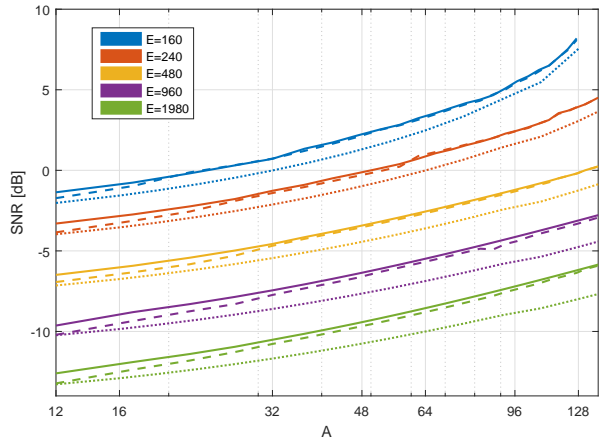}
\caption{SNR required under CK decoding (solid line) and CS decoding (dashed line) to reach BLER=$10^{-3}$ for various message lengths $A$ and codeword lengths $E$ over AWGN channel and BPSK modulation. List size is $L=8$. Corresponding PPV bound is represented by the dotted line.}
\label{fig:PDCCHAvsSNR}
\end{figure}

Finally, Figure~\ref{fig:PDCCHAvsSNR} provides the SNR needed to reach a BLER=$10^{-3}$ for several message lengths $A$ and codeword lengths $E$. 
The CK and CS decoders provide similar BLER performances for large message lengths $A$.
When the messages are composed of fewer bits ($A \leq 20$) , the performance of the CS decoder approaches the $\mathcal{O}\left(n^{-2}\right)$ approximation of PPV bound for each codeword length simulated.
\section{Conclusion} \label{sec:conc}
\balance

In this work, we described and compared three different incarnations of SCL-based decoders for DCA polar codes in 5G-NR, providing different trade-offs between performance, latency and complexity. 
For small list sizes, the CK decoder has a higher chance of early-termination with a similar BLER when compared to the CR decoder. 
As $L$ increases, early-terminations become increasingly rare and the performance gain brought by the CR and CS decoders is more substantial. 
If early termination is not envisaged, the CS decoder is preferable over the other two due to better performance and reduced decoding complexity. 
For PBCH codes in 5G-NR with 3 distributed CRC bits, CR and CS yield $0.08$ dB gain for $L=8$ and $0.2$ dB for $L=32$ with respect to CK.
PDCCH simulations show that, thanks to the 7 distributed CRC bits, early termination can occur more often, while the BLER gain brought by CR and CS is reduced. Moreover, when the message length $A$ is small, the gain using CS compared to CK increases up to $0.6$ dB and approaches the Polyanskiy-Poor-Verdù bound.


\begin{thebibliography}{10}

\bibitem{arikan}
E.~Ar{\i}kan,
\newblock ``Channel polarization: A method for constructing capacity-achieving
  codes for symmetric binary-input memoryless channels,''
\newblock {\em IEEE Transactions on Information Theory}, vol. 55, no. 7, pp.
  3051--3073, July 2009.

\bibitem{tal_list}
I.~Tal and A.~Vardy,
\newblock ``List decoding of polar codes,''
\newblock {\em IEEE Transactions on Information Theory}, vol. 61, no. 5, pp.
  2213--2226, May 2015.

\bibitem{CRC_aid}
K.~Niu and K.~Chen,
\newblock ``{CRC}-aided decoding of polar codes,''
\newblock {\em IEEE Communications Letters}, vol. 16, no. 10, pp. 1668--1671,
  October 2012.

\bibitem{Hashemi_5G}
S.~A. {Hashemi}, C.~{Condo}, F.~{Ercan}, and W.~J. {Gross},
\newblock ``On the performance of polar codes for {5G eMBB} control channel,''
\newblock in {\em 2017 51st Asilomar Conference on Signals, Systems, and
  Computers}, Oct 2017, pp. 1764--1768.

\bibitem{3GPP_TS}
$3^{\text{rd}}$ Generation Partnership Project~({3GPP}),
\newblock ``Technical specification group radio access network,''
\newblock {\em 3GPP TS 38.212 V.15.0.0}, 2017.

\bibitem{hui_dca}
D.~{Hui}, M.~{Breschel}, and Y.~{Blankenship},
\newblock ``Interleaved {CRC} for polar codes,''
\newblock in {\em IEEE Vehicular Technology Conference (VTC Spring)}, Porto,
  Portugal, June 2018.

\bibitem{dist_CRC_aid}
J.~Chen, Y~Chen, K.~Jayasinghe, D.~Du, and J.~Tan,
\newblock ``Distributing {CRC} bits to aid polar decoding,''
\newblock in {\em IEEE Global Communications Conference (GLOBECOM)}, Singapore,
  December 2017.

\bibitem{Blind_ET}
C.~{Condo}, S.~A. {Hashemi}, A.~{Ardakani}, F.~{Ercan}, and W.~J. {Gross},
\newblock ``Design and implementation of a polar codes blind detection
  scheme,''
\newblock {\em IEEE Transactions on Circuits and Systems II: Express Briefs},
  vol. 66, no. 6, pp. 943--947, June 2019.

\bibitem{PC_conc}
T.~Wang, D.~Qu, and T.~Jiang,
\newblock ``Parity-check-concatenated polar codes,''
\newblock {\em IEEE Communications Letters}, vol. 20, no. 12, pp. 2342--2345,
  December 2016.

\bibitem{design5G}
V.~Bioglio, C.~Condo, and I.~Land,
\newblock ``Design of polar codes in {5G} new radio,''
\newblock {\em CoRR}, vol. abs/1804.04389, 2018.

\bibitem{SpectreMatlab}
T.~{Erseghe},
\newblock ``Coding in the finite-blocklength regime: Bounds based on laplace
  integrals and their asymptotic approximations,''
\newblock {\em IEEE Transactions on Information Theory}, vol. 62, no. 12, pp.
  6854--6883, Dec 2016.

\end{thebibliography}
\end{document}